\documentclass[12pt]{article}
\usepackage{graphicx}
\usepackage{amsmath}

\begin{document}
\date{October 24, 2005}
\title{\begin{flushright}{\small IFUM-851-FT}\end{flushright}
Fermion masses in E$_{6}$ grand unification with family permutation
symmetries }
\author{{\small Francesco Caravaglios and Stefano Morisi} \\
{\small Dipartimento di Fisica, Universit\`{a} di Milano, Via Celoria 16,
I-20133 Milano, Italy} \\
{\small and}\\
{\small INFN\ sezione di Milano}}
\maketitle

\begin{abstract}
Neutrino masses have been extensively studied in the context of discrete
symmetries. A family permutation symmetry easily explains \ all neutrino
oscillations and their mixing angles. However it is not obvious the
embedding of such symmetry in a grand unified theory where all fermions live
in the same representation. We show that it is possible to embed such a
horizontal discrete symmetry in a grand unified theory. We discuss an E$_{6}$
\ GUT\ model as in \cite{Stech:2003sb}, but differently from 
\cite{Stech:2003sb}, we take the SM Higgs doublet in the 351$^\prime$ 
to have (+3,+5) charges with
respect the additional $U(1)_r\times U(1)_t$. This choice makes
possible two different horizontal symmetry breaking patterns and a
distinction between neutrinos and the rest of matter  fermions.
\end{abstract}

\section*{A grand unified model for fermion masses}

In this paper we will study a E$_{6}$ unified gauge group
\cite{Stech:2003sb,Maekawa:2003bb,Slansky:1981yr} that could explain
at the same time, neutrino, lepton and quark masses. In addition to the
gauge symmetry group we will require a permutation symmetry
\cite{Lam:2005va,carav2} of the three
fermion families. An horizontal permutation symmetry means that the three
 families can be considered identical in the sense of third 
quantization\cite{carav1}.
The breaking of this S$_{3}$ symmetry into a S$_{2}$
symmetry exchanging the $\nu_{\mu }$ \ $\nu_{\tau }$ \ neutrinos  can
easily explain large solar and atmospheric mixing angles\cite{exp1}. 
However this
permutation symmetry does not directly apply to the quark and lepton
 masses,
in which case both  S$_{3}$  and S$_2$ are strongly broken. The mass ratio
 of the muon and the
electron is roughly 200. The issue is how to embed \ a neutrino mass
matrix that is S$_{2}$ symmetric and with just a small breaking mass term of
the full S$_{3}$ group, in the same unified group with lepton and quarks
where the breaking of S$_{3}$ and S$_{2}$ is \ instead strong. The idea is
the following. Yukawa couplings for quark and leptons strongly break the
permutation symmetry, thus we expect that they are proportional to the
v.e.v. of some scalar fields responsible of the S$_{3}$ spontaneous symmetry
breaking. Probably yukawa couplings in the neutrino sectors are not
proportional to the same fields in order to explain this difference. This
possibility is not obvious when we are dealing with gauge groups with all
fermions living in the same irreducible representation. In SU(5) the Yukawa
couplings can be written in the following way 
\begin{equation}
L=g_{u}\,T^{\alpha \beta }\,T^{\gamma \delta }\,H^{\sigma }\,\varepsilon
_{\alpha \beta \gamma \delta \sigma }+g_{d}\,T^{\alpha \beta }\text{ }%
F_{\alpha }\,\bar{H}_{\beta }+g_{v}\,\ F_{\alpha }\,\nu_{R}\,H^{\alpha
}+M\,\nu_{R}^{t}\,\nu_{R}   \label{eq:1}
\end{equation}
where $T^{\alpha \beta },F_{\alpha }$ are weyl fermions belonging to the 10
and \={5} of SU(5). \ $\nu_{R}\,$\ is the SU(5)\ singlet right-handed
neutrino and M is a majorana mass term. In the lagrangian above, Yukawa
interactions are distinct, but we want to understand why these are
proportional to the v.e.v. of different scalars. If we embed SU(5) into
SO(10) we have an additional U(1) gauge group that commutes with the full
SU(5). The U(1) charges for the representation above are $H(+q),\bar{H}(-q),$
$T(-1)$ $F(+3)$ and $\nu_{R}(-5)$. From these charges we derive that \ each
mass operators have U$_{r}$(1) charges 
\begin{equation*}
\begin{tabular}{ll}
\hline
SU(5) mass operator & U$_{r}$(1) \\ 
\hline
&\\
$T^{\alpha \beta }\text{ }F_{\alpha }$ & +2 \\ 
$F_{\alpha }\,\nu_{R}$ & -2 \\ 
$\nu_{R}^{t}\,\nu_{R}$ & -10 \\ 
$T^{\alpha \beta }\,T^{\gamma \delta }$ & -2\\
\hline
\end{tabular}
\end{equation*}
We observe that the mass operators $T^{\alpha \beta }\,T^{\gamma \delta }$
and $F_{\alpha }\,\nu_{R}$ have the same charges, thus we expect that the
same SU(5) singlet is at the origin of their Yukawa interaction. While
neutrinos have an approximate S$_{3}$ symmetry \cite{carav2}, 
the same symmetry 
 is not observed in the
 up sector. If we embed SU(5) into E$_{6}$ we have an additional U$_{t}$%
(1) and the table above becomes

\begin{equation*}
\begin{tabular}{lll}
\hline
SU(5) mass operator & U$_{r}$(1) & U$_{t}$(1) \\ 
\hline
&&\\
$T^{\alpha \beta }\text{ }F_{\alpha }$ & +2 & +2 \\ 
$F_{\alpha }\,\nu_{R}$ & -2 & +2 \\ 
$\nu_{R}^{t}\,\nu_{R}$ & -10 & +2 \\ 
$T^{\alpha \beta }\,T^{\gamma \delta }$ & -2 & +2 \\ 
$F_{\alpha }\,x_{L}$ & +3 & +5 \\ 
$\nu_{R}^{t}\,x_{L}$ & -5 & +5 \\ 
$x_{L}^{t}\,x_{L}$ & 0 & +8\\
\hline
\end{tabular}
\end{equation*}
The advantage here is that we have two right-handed neutrinos $\nu_{R}$ 
and $%
x_{L},$ and the Dirac mass operator $F_{\alpha }\,x_{L}$  has different
quantum numbers from all the others and in particular is different from $%
T^{\alpha \beta }\,T^{\gamma \delta }$ giving mass to the \ up sector. Thus
we explore the possibility that the fundamental lagrangian has a E$_{6}$
unifying gauge symmetry times a S$_{3}$ permutation symmetry of the three
fermion families that belong to the 27 of E$_{6}$. As already said the 27
contains two standard model singlets that will play the role of
 right-handed
neutrinos. Now we have to choose the representation for the Higgs
 SU(2)$_{W}$
doublet. We prefer to keep just one Higgs doublet , that will give
 mass both
for the up and the down sector. This is because we want to have the
 Standard
Model with just one higgs at \ the weak scale where the FCNC\ are strongly
suppressed due to the GIM\ mechanism. So we choose the Higgs to be also a
 S$%
_{3}$ singlet. Now we have to decide to which E$_{6}$ representation we have
to assign the Higgs doublet. The Yukawa interaction for fermions at the tree
level can be 
\begin{equation}
27_{i}^{\alpha }\,\ 27_{i}^{\beta }\,351_{\alpha \beta }^{\prime }\,\ \ \ \
\ \ \ \ \ \ \ \ \ \ \ i=\text{1,2,3 family index}   \label{eq:2}
\end{equation}
where the 351$^{\prime }$ is symmetric under the exchange of $\alpha $ and $%
\beta $ the gauge symmetry indices. The reason why we put the Higgs doublet
in the 351$^{\prime }$ is that it contains a SU(2)$_{W}$ doublet with the
(-3,-5) charges with respect the U$_{r}$(1)$\times $U$_{t}$(1). Also the 351
contains a Higgs doublet with same charges, but this would give zero in (\ref
{eq:2}). At the tree level of the fundamental high energy lagrangian, we
have just one Yukawa interaction that comes from (\ref{eq:2}), since this is
the unique U$_{r}$(1)$\times $U$_{t}$(1) gauge invariant \ operator 
\begin{equation*}
g\,\ x_{iL}^{t}\,\ \nu_{iL}\,h_{0}
\end{equation*}
As already mentioned the higgs is a S$_{3}$ singlet and does not carry any
family index, while fermions are assigned to the 27$_{i}$ and $i$ runs over
the three fermion families. In this way we have been able to accommodate a
Dirac mass proportional to the identity matrix as proposed in 
\cite{carav2}. So, before the E$_{6}$ symmetry breaking, quark and charged
lepton yukawa couplings are zero, since they do not form a gauge invariant
operator with the Standard Model Higgs. The up quark yukawa operator (in a
SU(5) notation) is $T^{\alpha \beta }\,T^{\gamma \delta }H^{\sigma
}\,\varepsilon _{\alpha \beta \gamma \delta \sigma }$ and its charges are $%
(+5,+3)$. We need a SU(5) singlet with opposite U$_{r}$(1)$\times $U$_{t}$%
(1) to make an invariant operator. At first sight such a singlet is
contained both in the 78 and in the 650, it has the correct U(1) charges.
But we will see after that  in order to give a Yukawa coupling to the up 
quarks we have to write
an interaction 
\begin{equation}
27^{\alpha }\,27^{\beta }351_{\gamma \sigma }^{\prime }\Sigma _{\alpha \beta
}^{\gamma \sigma }   \label{eq:4}
\end{equation}
where $\Sigma _{\alpha \beta
}^{\gamma \sigma } $ is the irrep 2430.
\section*{The breaking $E_6\supset SU(2)\times SU(6)\supset
 U(1)_x\times SU(5)$}
The breaking $E_6\supset SU(2)\times SU(6)$ can be achieved after the 
irrep 650 takes a vev.  If $Q_{r}$ and $Q_{t}$ are the U(1)\ charges
corresponding to the embedding E$_{6}\supset Q_{t}\times SO(10)\supset
Q_{t}\times Q_{r}$ $\times SU(5)$ while Q$_{T_{3}}$ and Q$^{\prime }$
correspond to the embedding E$_{6}\supset SU(2)\times SU(6)\supset
Q_{T_{3}}\times SU(6)\supset $%
\par $Q_{T_{3}}\times $Q$^{\prime }\times SU(5)$, then these charges 
 are related as follows
\begin{equation*}
\begin{array}{l}
Q_{r}=-\frac{1}{2}Q^{\prime }-\frac{5}{2}Q_{T_{3}} \\ 
Q_{t}=\frac{1}{2}Q^{\prime }-\frac{3}{2}Q_{T_{3}}
\end{array}
\Rightarrow 
\begin{array}{l}
Q^{\prime }=-\frac{3}{4}Q_{r}+\frac{5}{4}Q_{t} \\ 
Q_{T_{3}}=-\frac{1}{4}Q_{r}-\frac{1}{4}Q_{t}
\end{array}
.
\end{equation*}
All the quantum numbers are reported in table \ref{tab3}. The branching rule
for the 27 is 27=(2,\={6})+(1,15) where the branching rule for \={6} in SU(5)%
$\times $U(1)$^{\prime }$ is \={6}=1(5)+\={5}(-1) and for 15 is
15=5(-4)+10(2). That means that we can take two tensors that contain all
fermions $T_{(\alpha \beta )}$ and $R_{a}^{\alpha }$ (see table \ref{tab1}). 
\begin{table}
\begin{equation*}
T_{(\alpha \beta )}=\left( 
\begin{tabular}{llllll}
0 & $\bar{D}_{1}$ & $\bar{D}_{2}$ & $\bar{D}_{3}$ & -$N_{L}$ & $E_{L}$ \\ 
-$\bar{D}_{1}$ & 0 & $u_{R3}^{c}$ & -$u_{R2}^{c}$ & -$d_{L1}$ & $u_{L1}$ \\ 
-$\bar{D}_{2}$ & -$u_{R3}^{c}$ & 0 & $u_{R1}^{c}$ & -$d_{L2}$ & $u_{L2}$ \\ 
-$\bar{D}_{3}$ & $u_{R2}^{c}$ & -$u_{R1}^{c}$ & 0 & -$d_{L3}$ & $u_{L3}$ \\ 
$N_{L}$ & $d_{L1}$ & $d_{L2}$ & $d_{L3}$ & 0 & $e_{R}^{c}$ \\ 
-$E_{L}$ & -$u_{L1}$ & -$u_{L2}$ & -$u_{L3}$ & -$e_{R}^{c}$ & 0
\end{tabular}
\right) _{\alpha \beta }
\end{equation*}
\begin{equation*}
R_{a}^{\alpha }=\left( 
\begin{tabular}{llllll}
$\nu_{R1}^{c}$ & $D_{1}$ & $D_{2}$ & $D_{3}$ & $\bar{N}_{L}$ & $\bar{E}_{L}$
\\ 
$x_{L}$ & $d_{R1}^{c}$ & $d_{R2}^{c}$ & $d_{R3}^{c}$ & $\nu_{L}$ & $e_{L}$%
\end{tabular}
\right) _{a\alpha }
\end{equation*}
\caption{Branching of \textbf{27 } in SU(2)$\times $SU(6): 27= (2,$\bar{6}$%
)+(1,15)=$T_{(\alpha \beta )}+R_{a}^{\alpha }$}
\label{tab1}
\end{table}
We use the following convention for the tensors, upper indices and lower
indices differs for a complex conjugation, i.e. $R_{a}^{\alpha }\,R_{\alpha
}^{a}$ is a SU(2)$\times $SU(6) invariant. Latin indices $a,b,c,...=1,2$ are
the SU(2)\ indices, while greek indices $\alpha ,\beta ,\gamma ,...=1,6$ \
are the SU(6) indices.$i,j,k=1,2,3$ are family indices. Sometimes, \
whenever confusion is possible, underlined indices \underline{$\alpha $} \
mean E$_{6}$ indices and run from 1 to 27. We also use the convention that a
tensor $T_{(\alpha \beta \gamma )}$ is antisymmetric under permutations of
the indices within the parenthesis () while it is symmetric if indices are
within [] parenthesis as in $T_{[\alpha \beta \gamma ]}.$ The higgs doublet
is in the $351^{\prime }=(1,15)+(3,21)+(2,\overline{84})+(1,105^{\prime }),$
and more precisely it lives in the (3,21), i.e. we can consider a tensor $%
K_{[\alpha \beta ]}^{[ab]}$ \ of SU(2)$\times $SU(6) that includes the
standard Higgs doublet. The higgs doublet has charges Q$_{r}=-3$ and Q$%
_{t}=-5$ which corresponds to the U(1) charges Q$%
_{T_{3}}=2$ and Q$^{\prime }=4$ with respect U$_{T_{3}}$(1)$\times $U(1)$%
^{\prime }$.
\begin{table}
\begin{equation*}
\begin{array}{l}
\hline
{\small ~~~\mbox{SM}~~~~~Q_{em}~~~~Y~~~~~I_{3}^{w}~~~Q_{r}~~~~Q_{t}~~~~Q^{%
\prime }~~~~Q_{T_{3}}} \\ \hline
~ \\ 
\left[ 
\begin{tabular}{cccccccc}
${\bar{D}_{1}}$ & $-{\frac{1}{3}}$ & $-{\frac{1}{3}}$ & $0$ & $2$ & $-2$ & $%
-4$ & $0$ \\ 
$\bar{D}_{2}$ & $-{\frac{1}{3}}$ & $-{\frac{1}{3}}$ & $0$ & $2$ & $-2$ & $-4$
& $0$ \\ 
${\ u_{R3}^{c}}$ & $-{\frac{2}{3}}$ & $-{\frac{2}{3}}$ & $0$ & $-1$ & $1$ & $%
2$ & $0$ \\ 
$\bar{D}_{3}$ & $-{\frac{1}{3}}$ & $-{\frac{1}{3}}$ & $0$ & $2$ & $-2$ & $-4$
& $0$ \\ 
$u_{R2}^{c}$ & $-{\frac{2}{3}}$ & $-{\frac{2}{3}}$ & $0$ & $-1$ & $1$ & $2$
& $0$ \\ 
$u_{R1}^{c}$ & $-{\frac{2}{3}}$ & $-{\frac{2}{3}}$ & $0$ & $-1$ & $1$ & $2$
& $0$ \\ 
$N_{L}$ & $0$ & ${\frac{1}{2}}$ & $-{\frac{1}{2}}$ & $2$ & $-2$ & $-4$ & $0$
\\ 
$d_{L1}$ & $-{\frac{1}{3}}$ & ${\frac{1}{6}}$ & $-{\frac{1}{2}}$ & $-1$ & $1$
& $2$ & $0$ \\ 
$d_{L2}$ & $-{\frac{1}{3}}$ & ${\frac{1}{6}}$ & $-{\frac{1}{2}}$ & $-1$ & $1$
& $2$ & $0$ \\ 
$d_{L3}$ & $-{\frac{1}{3}}$ & ${\frac{1}{6}}$ & $-{\frac{1}{2}}$ & $-1$ & $1$
& $2$ & $0$ \\ 
${El}$ & $1$ & ${\frac{1}{2}}$ & ${\frac{1}{2}}$ & $2$ & $-2$ & $-4$ & $0$
\\ 
$u_{L1}$ & ${\ \frac{2}{3}}$ & ${\frac{1}{6}}$ & ${\frac{1}{2}}$ & $-1$ & $1$
& $2$ & $0$ \\ 
${u_{L2}}$ & ${\frac{2}{3}}$ & ${\frac{1}{6}}$ & ${\frac{1}{2}}$ & $-1$ & $1$
& $2$ & $0$ \\ 
${u_{L3}}$ & ${\frac{2}{3}}$ & ${\frac{1}{6}}$ & ${\frac{1}{2}}$ & $-1$ & $1$
& $2$ & $0$ \\ 
$e_{R}^{c}$ & $1$ & $1$ & $0$ & $-1$ & $1$ & $2$ & $0$ \\ 
${\nu _{R}^{c}}$ & $0$ & $0$ & $0$ & $-5$ & $1$ & $5$ & $1$ \\ 
${D_{1}}$ & ${\frac{1}{3}}$ & ${\frac{1}{3}}$ & $0$ & $-2$ & $-2$ & $-1$ & $%
1 $ \\ 
${D_{2}}$ & ${\frac{1}{3}}$ & ${\frac{1}{3}}$ & $0$ & $-2$ & $-2$ & $-1$ & $%
1 $ \\ 
${D_{3}}$ & ${\frac{1}{3}}$ & ${\frac{1}{3}}$ & $0$ & $-2$ & $-2$ & $-1$ & $%
1 $ \\ 
${\bar{N}_{L}}$ & $0$ & $-{\frac{1}{2}}$ & ${\frac{1}{2}}$ & $-2$ & $-2$ & $%
-1$ & $1$ \\ 
${\bar{E}_{L}}$ & $-1$ & $-{\frac{1}{2}}$ & $-{\frac{1}{2}}$ & $-2$ & $-2$ & 
$-1$ & $1$ \\ 
${x_{L}}$ & $0$ & $0$ & $0$ & $0$ & $4$ & $5$ & $-1$ \\ 
${d_{R1}^{c}}$ & ${\frac{1}{3}}$ & ${\frac{1}{3}}$ & $0$ & $3$ & $1$ & $-1$
& $-1$ \\ 
${d_{R2}^{c}}$ & ${\frac{1}{3}}$ & ${\frac{1}{3}}$ & $0$ & $3$ & $1$ & $-1$
& $-1$ \\ 
${d_{R3}^{c}}$ & ${\frac{1}{3}}$ & ${\frac{1}{3}}$ & $0$ & $3$ & $1$ & $-1$
& $-1$ \\ 
${\nu _{L}}$ & $0$ & $-{\frac{1}{2}}$ & ${\frac{1}{2}}$ & $3$ & $1$ & $-1$ & 
$-1$ \\ 
${e_{L}}$ & $-1$ & $-{\frac{1}{2}}$ & $-{\frac{1}{2}}$ & $3$ & $1$ & $-1$ & $%
-1$%
\end{tabular}
\right]
\end{array}
\end{equation*}
\caption{Quantum numbers of the irrep {\bf 27}} \label{tab3}
\end{table}

 The only representation with correct charges is (3,21). The
usual two components vector of the Higgs doublet are written in terms of the 
$K$ components as follows $(h_{0},h^{-})=(K_{[15]}^{[22]},K_{[16]}^{[22]}).$
The yukawa interaction ($i=1,2,3$ is the family index) 
\begin{equation}
L=27_{i}^{\underline{\alpha }}\,\ 27_{i}^{\underline{\beta }}\,\ 351_{%
\underline{\alpha \beta }}^{\prime }=R_{ai}^{\alpha }R_{bi}^{\beta }\text{ }%
K_{[\alpha \beta ]}^{[ab]}=R_{2i}^{1}R_{2i}^{5}\text{ }K_{[15]}^{[22]}=%
x_{Li} \nu_{Li}h_{0}    \label{eq:3}
\end{equation}
Thus there is only one Yukawa interaction in the fundamental E$_{6}$
symmetric and renormalizable lagrangian this is the Dirac neutrino mass . (%
\ref{eq:3}) does not introduce any mass neither  for quarks nor for charged 
leptons.
This result is important since a Yukawa interaction $u_{R\,\ i}^{c}\,\
u_{L\,\ i}\,h_{0}$ (symmetric under S$_{3}$ family permutations) would give $%
m_{\text{top}}=m_{\text{charm}}=m_{\text{up}}$ that is clearly unacceptable.
Quark and leptons can take a yukawa interaction only after the S$_{3}$
symmetry is broken. We expect they arise through higher order operators in
effective lagrangian approach after the spontaneous symmetry breaking of E$%
_{6}\times $S$_{3}$. This result is automatic because we have chosen the
Higgs doublet to  be in the 351$^{\prime},$ with $U(1)_r\times U(1)_t$
 charges (-3,-5).

\begin{table}[h]
\begin{equation*}
\begin{tabular}{lll}
\hline
SU(5) mass operator & U$_{r}$(1) & U$_{t}$(1) \\ 
\hline
&&\\
$\bar{D}^{\alpha }x_{L}H_{\alpha }^{\ast }\qquad\qquad$ & 5 & 7 \\ 
$x_{L}^{t}\,x_{L}$ & 0 & 8 \\ 
$\nu_{R}^{t}\,x_{L}$ & -5 & 5 \\ 
$F_{\alpha }\,x_{L}H^{\alpha }$ & 0 & 0 \\ 
$T^{\alpha \beta }\,T^{\gamma \delta }H^{\sigma } 
\epsilon_{\alpha \beta \gamma \delta \sigma}$& -5 & -3 \\ 
$T^{\alpha \beta }\text{ }F_{\alpha }H_{\beta }^{\ast }$ & 5 & 7 \\ 
$\nu_{R}^{t}\,\nu_{R}$ & -10 & 2 \\ 
$F_{\alpha }\,\nu_{R}H^{\alpha }$ & -5 & -3 \\ 
$D_{\alpha }\,x_{L}H^{\alpha }$ & -5 & -3 \\ 
$T^{\alpha \beta }\text{ }D_{\alpha }H_{\beta }^{\ast }$ & 0 & 4 \\ 
$D_{\alpha }\,\nu_{R}H^{\alpha }$ & -10 & -6 \\ 
$D_{\alpha }\bar{D}^{\alpha }$ & 0 & -4 \\ 
$F_{\alpha }\bar{D}^{\alpha }$ & 5 & -1 \\ 
$\bar{D}^{\alpha }\nu_{R}H_{\alpha }^{\ast }$ & 0 & 4\\
\hline
\end{tabular}
\end{equation*}
\caption{Mass terms in SU(5).}\label{tab3bis}
\end{table}

Now suppose that we want to give mass to the $D\bar{D}$, then we have to
break $U(1)_r\times U(1)_t$ in the direction (0,4), but in this case
also the $x_{L}x_{L}$ gets a mass and would be heavier than the $\nu_{R}$.
One could break the U(1) and give a mass slightly smaller than the $x_{L}$.
This is because we want the mixing between the $\nu_{L}$ and $x_{L}$ and a
too heavy $x_{L}$ would not allow this possibility. The situation looks
better if we first break $U(1)_r\times U(1)_t$ in the direction
(5,-5)\ giving a Dirac mass $\nu_{R}\,x_{L}$ (see table \ref{tab3bis}).
 This can be achieved through
the 351$^{\prime }$ that could come from the product $78\times 351$ (since
the Higgs doublet is already in the 351$^{\prime }$). Thus the breaking
could proceed as follows . First the 650 takes a v.e.v. and breaks the E$%
_{6}\supset $SU(2)$\times $SU(6) ,after that the 78 could break SU(2)$\times 
$SU(6)$\supset $U(1)$\times $SU(6) and the 351 breaks U(1)$\times $SU(6)$%
\supset $SU(5)$\times $U(1) , then the tensor 351$^{\prime }\in $78$\times $%
351 \ can give dirac mass $\nu_{R}\,x_{L}$ through the operator $\
27^{\alpha }27^{\beta }351_{\alpha \gamma }78_{\beta }^{\gamma }$. So we are
left with SU(5)$\times $U(1)$_{x}$ and the charges with respect this new U(1)%
$_{x}$ are in the table \ref{tb}.

Thus we see that the breaking of the remaining U(1) will give mass to all
fermions and the necessary Yukawa couplings. Note that if the breaking
occurs through a field with charges (0,4), il will appear as a second power
\ for the up-sector and at the third power in the down sector, since they
have respectively charges 8 and 12. This is compatible with the fact that
the up-quark sector is heavier than the down sector. Concerning the S$_{3}$
symmetry breaking, we can imagine in first approximation that the \ 351 is
responsible for the S$_{3}$ breaking. Let us discuss first the neutrino
sector. We can have the following lagrangian 
\begin{equation*}
L=27_{i}^{\underline{\alpha }}\,\ 27_{j}^{\underline{\beta }}\,\ 351_{k\,\ 
\underline{\alpha \gamma }}78_{\beta }^{\gamma }+27_{i}^{\underline{\alpha }%
}\,\ 27_{i}^{\underline{\beta }}\,\ 351_{i\,\ \underline{\alpha \gamma }%
}78_{\beta }^{\gamma }
\end{equation*}

Light left handed neutrinos will mix with the heavy dirac neutrino through
the mass matrix \ (i and j family index) 
\begin{equation*}
M=\left( 
\begin{tabular}{lll}
$\nu_{L}$ & $\nu_{R}^{c}$ & $x_{L}$%
\end{tabular}
\right) ^{i}\left( 
\begin{tabular}{lll}
0 & 0 & $m\,\ \delta _{ij}$ \\ 
0 & $m_{v}\,\ \delta _{ij}$ & $M_{ij}$ \\ 
$m\,\ \delta _{ij}$ & $M_{ij}$ & $m_{x}\,\ \delta _{ij}$%
\end{tabular}
\right) \left( 
\begin{tabular}{l}
$\nu_{L}$ \\ 
$\nu_{R}^{c}$ \\ 
$x_{L}$%
\end{tabular}
\right) ^{j}
\end{equation*}
that will give the following mass matrix elements for the left-handed 
neutrinos $(M>>m,m_\nu,m_x)$ 
\begin{equation*}
m^{ij}_{\mathrm{Light}}=m_{v}\left( \frac{m^{2}}{M^{2}}\right) _{ij}
\end{equation*}
If $m$ and $m_{v}$ are proportional to the identity matrix and $m_{v}\sim
m_{x}$ (since from table 3 we see that both have U(1)$_{x}$ (8)) only $M$
will be responsible for S$_{3}$ breaking in the neutrino sector. Then we
have to deal with the up and down quark S$_{3}$ breaking. From tensor
analysis we see that to introduce the yukawa for up quarks \ we need the
2430 acquiring a v.e.v. 
\begin{table}
\begin{center}
\begin{tabular}{lll}
\hline
SU(5) mass operator & U$_{x}$(1) &  \\ 
\hline
&&\\
$\bar{D}^{\alpha }x_{L}H_{\alpha }^{\ast }$ & 12 &  \\ 
$x_{L}^{t}\,x_{L}$ & 8 &  \\ 
$\nu_{R}^{t}\,x_{L}$ & 0 &  \\ 
$F_{\alpha }\,x_{L}H^{\alpha }$ & 0 &  \\ 
$T^{\alpha \beta }\,T^{\gamma \delta }H^{\sigma }$ & -8 &  \\ 
$T^{\alpha \beta }\text{ }F_{\alpha }H_{\beta }^{\ast }$ & 12 &  \\ 
$\nu_{R}^{t}\,\nu_{R}$ & -8 &  \\ 
$F_{\alpha }\,\nu_{R}H^{\alpha }$ & -8 &  \\ 
$D_{\alpha }\,x_{L}H^{\alpha }$ & -8 &  \\ 
$T^{\alpha \beta }\text{ }D_{\alpha }H_{\beta }^{\ast }$ & 4 &  \\ 
$D_{\alpha }\,\nu_{R}H^{\alpha }$ & -16 &  \\ 
$D_{\alpha }\bar{D}^{\alpha }$ & -4 &  \\ 
$F_{\alpha }\bar{D}^{\alpha }$ & 4 &  \\ 
$\bar{D}^{\alpha }\nu_{R}H_{\alpha }^{\ast }$ & 4 &\\
\hline 
\end{tabular}
\end{center}\caption{U(1)$_x$ charges} \label{tb}
\end{table}
 In fact even if the 2430 can be obtained as the
product of the 351$\times $27$\times $27 , and even if  the SU(5) singlet 
with U(1)$_{x}$
charge +8 in the 2430 can  be obtained as the tensor product 351$\times $27%
$\times $27 where \ both the 351 and the 27 are SU(5) singlets, in order to
turn on the \ up quark yukawa coupling we need to add the 2430 to the list
of representations of the scalar fields.
The reason is the following.

 The top Yukawa interaction is \ $T^{\alpha
\beta }\,\ T^{\gamma \delta }H^{\sigma }\varepsilon _{\alpha \beta \gamma
\delta \sigma }.$ If the field H is in the 351$^\prime$, and 
 it has charges (-3,-5), then indices with these charges correspond to the $%
351_{22,26}^{\prime }$ (see table 3 to see the U(1) charges relative to \
22nd and 26th rows of the \textbf{27}).

\bigskip The up quark yukawa coupling, written in E$_{6}$ notation becomes 
\begin{equation*}
\left(u_{L}^{1}u_{R}^{1c}+u_{L}^{2}u_{R}^{2c}+u_{L}^{3}u_{R}^{3c}\right)h=
\left(27^{12}27^{6}+27^{13}27^{5}+27^{14}27^{3}\right)351_{22,26}^{\prime }.
\end{equation*}
In order to make an E$_{6}$ invariant, we \ need to multiply the operator
above with few scalar representations that acquire some v.e.v. At the same
time, these v.e.v.'s must be \ neutral under the four U(1) charges of the
cartan subalgebra of SU(3)$\times $SU(2)$\times $U(1) (i.e. they must be
standard model singlets). A tensor, that is a standard model singlet, either
must simultaneously contain all indices 12,6,26 (see table \ref{tab3})
 or none of
them. The minimal representation containing $T_{12,6}^{26}$ is the 1728 of E$%
_{6}.$ We prefer to add the 2430$_{12,6}^{22,26}$ and make directly the
invariant 
\begin{equation*}
\left(27^{12}27^{6}\, 2430_{12,6}^{22,26}+ 27^{13}27^{5}\,
2430_{13,5}^{22,26}+ 27^{14}27^{3}\,
2430_{14,3}^{22,26}\right)351_{22,26}^{\prime }
\end{equation*}
since it could be helpful to explain the top-bottom hierarchy.

This is interesting since the 2430 could explain while the S$_{3}$ breaking
in the quark sector is not the same as in the neutrino sector. The
minimization of an effective potential could give a vev for the 2430$^{i}$
not proportional to the 351$^{i}$. Namely it could give a reversed
hierarchy. In addition we expect the vev in the 2430 to be proportional to
the 351$\times $27$\times $27 since its charge is (5,3) that is the sum of
(5,-5)+(0,4)+(0,4). Similar arguments lead us to introduce the 1728 to turn
on the yukawa coupling for the down quarks. We expect it could take a vev
proportional to 351$\times $27$\times $27$\times $27. The additional 27 can
be used to explain the mass ratio between the top and the bottom quarks.
Note that the up quark yukawa interaction arises from just one yukawa
interaction 
\begin{equation*}
27^{\alpha }\,\ 27^{\beta }\,\ 351_{\gamma \delta }^{\prime }\,\
2430_{\alpha \beta }^{\gamma \delta }
\end{equation*}
while the down quark yukawa we have two distinct operators available 
\begin{equation*}
k_{1}\,\ 27^{\alpha }\,\ 27^{\beta }\,\ 351^{\prime \ast \,\gamma \delta
}\,\ 1728_{\delta \beta }^{\ast \rho }\,\ \ \varepsilon _{\alpha \gamma \rho
}+k_{2}\,\ 27^{\alpha }\,\ 27^{\beta }\,\ 351^{\prime \ast \,\gamma \delta
}\,\ 1728_{\delta \alpha }^{\ast \rho }\,\ \varepsilon _{\beta \gamma \rho }.
\end{equation*}
Their distinction arises because there are two distinct way to contract the
indices of the E$_{6}$ invariant tensor $\varepsilon _{\beta \gamma \rho }$.
While the up matrix must be left-right symmetric the down quark matrix could
be not symmetric due to these two distinct contraction available\footnote{%
This result is similar to SU(5), where the up quark matrix must be
left-right symmetric, while the down matrix can be not symmetric.}. Namely 
\begin{eqnarray*}
&&g_{0}\,\ 27_{i}^{\alpha }\,\ 27_{i}^{\beta }\,\ 351^{\prime \ast \,\gamma
\delta }\,\ 1728_{i\,\delta \beta }^{\ast \rho }\,\ \varepsilon _{\alpha
\gamma \rho }+g_{1}\,\ 27_{i}^{\alpha }\,\ 27_{j}^{\beta }\,\ 351^{\prime
\ast \,\gamma \delta }\,\ 1728_{j\,\delta \alpha }^{\ast \rho }\,\
\varepsilon _{\beta \gamma \rho } \\
&&+g_{2}\,27_{i}^{\alpha }\,\ 27_{j}^{\beta }\,351^{\prime \ast \,\gamma
\delta }\,\ 1728_{j\,\delta \beta }^{\ast \rho }\,\ \varepsilon _{\alpha
\gamma \rho }+g_{2}\,27_{i}^{\alpha }\,\ 27_{j}^{\beta }\,351^{\prime \ast
 \,\gamma \delta }\,\ 1728_{k\,\delta \beta }^{\ast \rho }\,\
 \varepsilon _{\alpha\gamma \rho }
\end{eqnarray*}
after the 1728$^{\ast i}$ takes a v.e.v. we arrive at the following yukawa
for the down quark 
\begin{eqnarray*}
\sum_{ij}\,(g_{0}\,d_{L\,i}\,d_{R\,i}^{c}\,\ h\ \ \left\langle \overline{1728%
}_{i}\right\rangle \, &+&g_{1}\,d_{L\,i}\,d_{R\,j}^{c}\,\ h\,\ \
\left\langle \overline{1728}_{i}\right\rangle +g_{2}\,\ d_{L\,j}\,\
d_{R\,i}^{c}\,\ h\,\ \ \left\langle \overline{1728}_{i}\right\rangle )+ \\
&+&\sum_{i\neq j\neq k}g_{3}\,\ d_{L\,i}\,\ d_{R\,j}^{c}\,\ h\,\ \
\left\langle \overline{1728}_{k}\right\rangle
\end{eqnarray*}
from which we get the following mass matrix 
\begin{equation*}
M=\left( 
\begin{array}{ccc}
x_{1} & g_{1}x_{1}+g_{2}x_{2}+g_{3}x_{3} & g_{1}x_{1}+g_{3}x_{2}+g_{2}x_{3}
\\ 
g_{2}x_{1}+g_{1}x_{2}+g_{3}x_{3} & x_{2} & g_{3}x_{1}+g_{1}x_{2}+g_{2}x_{3}
\\ 
g_{2}x_{1}+g_{3}x_{2}+g_{1}x_{3} & g_{3}x_{1}+g_{2}x_{2}+g_{1}x_{3} & x_{3}
\end{array}
\right)
\end{equation*}
where $x_{i}=h~\langle \overline{1728}_{i}\rangle $ with $i=1,2,3$. The
latter fits the experimental data (the six quark masses and the three
angles, plus one phase in the CKM mixing matrix) with the following
parameters

\begin{center}
{\small 
\begin{tabular}{|l|c|}
\hline
&  down \\ \hline
$g_{1}$ & 0.401001 + i 0.0564 \\ \hline
$g_{2}$ & -0.00016 - i 0.000017 \\ \hline
$g_{3}$ & -0.0041 + i 0.0072 \\ \hline
$x_{1}$ & 0.0004 \\ \hline
$x_{2}$ & 0.1119 \\ \hline
$x_{3}$ & 2.1 \\ \hline
\end{tabular}
}
\end{center}

We observe that $x_{1},x_{2}\ll x_{3}$ and in the down sector $g_{1}\gg
g_{2} $ \ ( $g_{0}$ has been reabsorbed in the definition of x$_{i}$). The
up quark mass matrix is very close to the identity matrix and the left-right
symmetry is not strongly broken.\newline
We are now able to write the full lagrangian in E$_{6}$ notation \
responsible for all fermion masses

\begin{eqnarray*}
L &=&27_{i}^{\alpha }\,\ 27_{i}^{\beta }\,\ 351_{\gamma \delta }^{\prime
}+27_{i}^{\alpha }\,\ 27_{j}^{\beta }\,\ \left( 351_{\gamma \alpha }^{k}\,\
78_{\beta }^{\gamma }+351_{\gamma \beta }^{k}\,\ 78_{\alpha }^{\gamma
}\right) +27_{i}^{\alpha }\,\ 27_{i}^{\beta }\,\ \left( 351_{\gamma \alpha
}^{i}\,\ 78_{\beta }^{\gamma }+351_{\gamma \beta }^{i}\,\ 78_{\alpha
}^{\gamma }\right) \\
&&+27_{i}^{\alpha }\,\ 27_{i}^{\beta }\,\ 351_{\gamma \delta }^{\prime
}\,2430_{\alpha \beta }^{i\,\ \gamma \delta }\,+27_{i}^{\alpha }\,\
27_{i}^{\beta }\,\ 27_{k\,\alpha }^{\ast }\,\ 27_{k\,\beta }^{\ast
}+27_{i}^{\alpha }\,\ 27_{i}^{\beta }\,\ 351^{\prime \ast \,\ \gamma \delta
}\,\ 1728_{i\,\ \beta \delta }^{\rho }\,\varepsilon _{\alpha \gamma \rho } \\
&&+27_{i}^{\alpha }\,\ 27_{j}^{\beta }\,\ 351^{\prime \ast \,\ \gamma \delta
}\,\ 1728_{i\,\ \beta \delta }^{\rho }\,\varepsilon _{\alpha \gamma \rho
}+27_{i}^{\alpha }\,\ 27_{i}^{\beta }\,\ 27_{k\,}^{\,\gamma }\,\
27_{k\,}^{\,\delta }\,\ \ 351_{\gamma \alpha }^{k}\,\ 351_{\delta \beta
}^{k}+27_{i}^{\alpha }\,\ 27_{i}^{\beta }\,\ 27_{k\,}^{\gamma }\,\varepsilon
_{\alpha \beta \gamma }\ 
\end{eqnarray*}
and finally   (after all standard model singlets take a vev) 
we obtain the following lagrangian in $SU(2)\times SU(6)$ notation  
\begin{eqnarray*}
L &=&\nu_{R\,i}^{t}\,x_{L\,j}\left( (1,21)_{[11]}^{k}(3,1)^{[12]}+\delta
^{ij}(1,21)_{[11]}^{i}(3,1)^{[12]}\right) +u_{L}^{i}u_{R}^{ci}\ \ \
h_{0}\left( (3,35)_{[22]1}^{1i}-(3,35)_{[22]5}^{5i}\right) + \\
&&+\left( x_{Lj}\,N_{Li}+d_{Rj}^{c}d_{Li}+e_{Lj}e_{Ri}^{c}\right) \,\
h_{0}\,\ (4,6)_{i1}^{[222]}+\left(
x_{Li}\,N_{Li}+d_{Ri}^{c}d_{Li}+e_{Li}e_{Ri}^{c}\right) \,\ h_{0}\,\
(4,6)_{i1}^{[222]}+ \\
&&+\left( \bar{D}D+\bar{N}_{L}N_{L}+\bar{E}_{L}E_{L}\right) (2,\bar{6}%
)_{2}^{1k}+x_{L}^{i}x_{L}^{i}(\bar{2},6)_{1}^{2k}(\bar{2},6)_{1}^{2k}+%
x_{Li}\,\ \nu_{Li}\,\ \ h_{0}
\end{eqnarray*}
As already said the vev of 2430 and 1728 \ could show a reversed hierarchy
with respect the 351.

We can imagine the following gauge symmetry breaking pattern $%
E_{6}\rightarrow SU(2)\times SU(6)\rightarrow U(1)\times SU(5)\rightarrow
SU(5)$, and scalars take a vev \ following the hierarchy pattern\ $%
\left\langle 650\right\rangle >\left\langle 351\right\rangle \sim
\left\langle 78\right\rangle >\left\langle 27_{16}\right\rangle \sim
\left\langle 27_{22}\right\rangle $ \ and the $\left\langle
2430\right\rangle \sim \left\langle 351\right\rangle \,\ \left\langle
27\right\rangle ^{2}$ and $\left\langle 1728\right\rangle \sim \left\langle
351\right\rangle \,\ \left\langle 27\right\rangle ^{3}.$

\section*{A list of  all relevant masses and their origin in this
 E$_{6}$ breaking
pattern.}

\subsubsection*{$\mathbf{v}_{R\,}^{t}\mathbf{\,x}_{L}$}

This Dirac mass term appears when both the 351 and the 78 of E$_{6}$ take a
vev. E$_{6}$ has been broken into SU(2)$\times $SU(6) by the 650, and after
both the 351 and the 78 break SU(2)$\times $SU(6) into SU(5)$\times $U(1).
The 351$_{\alpha \beta }$ \ \ has charge (5,-5), but it is antisymmetric
under the exchange of the indices $\alpha \beta $. So we need to add the 78
to build a non zero invariant operator 
\begin{equation*}
\mathbf{27}_{i}^{\alpha }\,\ \mathbf{27}_{j}^{\beta }\,\ \left( 351_{\gamma
\alpha }^{k}\,\ 78_{\beta }^{\gamma }+351_{\gamma \beta }^{k}\,\ 78_{\alpha
}^{\gamma }\right) +\mathbf{27}_{i}^{\alpha }\,\ \mathbf{27}_{i}^{\beta }\,\
\left( 351_{\gamma \alpha }^{i}\,\ 78_{\beta }^{\gamma }+351_{\gamma \beta
}^{i}\,\ 78_{\alpha }^{\gamma }\right).
\end{equation*}
$i,j,k$ are family indices, and the 351$^{k}$ breaks the S$_{3}$ symmetry,
but this operator gives a mass only to the S$_{3}$ singlet component of the
dirac heavy neutrinos. In SU(2)$\times $SU(6) notation the operator above
can be written ($\lambda $ is a dimensionless constant) 
\begin{eqnarray*}
\mathbf{R}_{ai}^{\alpha }\mathbf{R}_{bj}^{\beta }(1,21)_{[\alpha \beta
]}^{k}\,(3,1)^{[ab]} &=&\mathbf{R}_{1i}^{1}\mathbf{R}%
_{2j}^{1}(1,21)_{[11]}^{k}(3,1)^{[12]} \\
&=&\mathbf{v}_{R\,i}^{t}\mathbf{\,x}_{L\,j}\left( \lambda
(1,21)_{[11]}^{k}(3,1)^{[12]}
+\delta^{ij}(1,21)_{[11]}^{i}(3,1)^{[12]}\right)
\end{eqnarray*}
where the (1,21) and (3,1) under SU(2)$\times $SU(6) are contained
respectively in the 351 and 78. We remind that latin letters run from 1 to 2
and are SU(2) indices while greek letters are SU(6)\ indices. This mass term
can be written as follows 
\begin{equation*}
M=c\left( 
\begin{array}{ccc}
x_{L1} & x_{L2} & x_{L3}
\end{array}
\right) \left( 
\begin{array}{ccc}
(1,21)_{[11]}^{1}/c+1 & 1 & 1 \\ 
1 & (1,21)_{[11]}^{2}/c+1 & 1 \\ 
1 & 1 & (1,21)_{[11]}^{3}/c+1
\end{array}
\right) \left( 
\begin{array}{c}
\nu_{R1}^{c} \\ 
\nu_{R2}^{c} \\ 
\nu_{R3}^{c}
\end{array}
\right)
\end{equation*}
and $c=\lambda \sum_{k}(1,21)_{[11]}^{k}(3,1)^{[12]}$ . In the limit $%
\lambda \rightarrow \infty $ the above matrix\cite{carav2} is diagonalized 
by the
tri-bimaximal matrix 
\begin{equation*}
\left( 
\begin{array}{ccc}
-2/\sqrt{6} & 1/\sqrt{3} & 0 \\ 
1/\sqrt{6} & 1/\sqrt{3} & -1/\sqrt{2} \\ 
1/\sqrt{6} & 1/\sqrt{3} & 1/\sqrt{2}
\end{array}
\right)
\end{equation*}
So this matrix could explain neutrino oscillations. To do that, we need to
mix these heavy neutrinos with the light left-handed neutrino and to give a
majorana mass to the $\nu_{R}^{c}\nu_{R}^{c}$.

\subsubsection*{$\mathbf{gv}_{L}^{i}\mathbf{x}_{L}^{i}\mathbf{h}$}

This mass arises directly from the tree level Yukawa interaction in the
fundamental E$_{6}$ invariant lagrangian. 
\begin{equation*}
27_{i}^{\alpha }\,\ 27_{i}^{\beta }\,\ 351_{\alpha \beta }^{\prime
}=R_{ai}^{\alpha }\,\ R_{bi}^{\beta }\,\ \ (3,21)_{[\alpha \beta
]}^{[ab]}=R_{2i}^{1}\,\ R_{2i}^{5}\,\ \ (3,21)_{[15]}^{[22]}=x_{Li}\,\
\nu_{Li}\,\ \ h_{0}
\end{equation*}
The Standard model higgs doublet $h$ is in the 351$^{\prime }$ and \ more
precisely in the component $\ (3,21)_{[15]}^{[22]}$ of SU(2)$\times $SU(6).
This dirac mass does not break S$_{3}$, in fact this mass matrix is
proportional to the identity matrix.

\subsubsection*{$\mathbf{v}_{R}^{c}\mathbf{v}_{R}^{c}$ and $\mathbf{x}_{L}%
\mathbf{x}_{L}$}

These majorana masses are necessary to break the lepton number and give mass
to the light neutrinos. They could arise from the the operator 
\begin{equation*}
\mathbf{27}_{i}^{\alpha }\mathbf{\,\ 27}_{i}^{\beta }\,\ 27_{k\,\alpha
}^{\ast }\,\ 27_{k\,\beta }^{\ast }=\mathbf{R}_{ai}^{\alpha }\mathbf{\,\ R}%
_{bi}^{\beta }\,\ (\bar{2},6)_{\alpha }^{ak}(\bar{2},6)_{\beta }^{bk}=%
\mathbf{R}_{2i}^{1}\mathbf{\,\ R}_{2i}^{1}\,\ (\bar{2},6)_{1}^{2k}(\bar{2}%
,6)_{1}^{2k}
\end{equation*}
or 
\begin{equation*}
\mathbf{27}_{i}^{\alpha }\mathbf{\,\ 27}_{i}^{\beta }\,\ 27_{k\,}^{\,\gamma
}\,\ 27_{k\,}^{\,\delta }\,\ \ 351_{\gamma \alpha }^{k}\,\ 351_{\delta \beta
}^{k}
\end{equation*}
These masses should not break the S$_{3}$ symmetry \ and should be
proportional to the identity matrix. That is the reason why family indices $%
i $ and $k$ are not the same for fermions and scalars in the operator above.

\subsubsection*{$\mathbf{u}_{L}\mathbf{u}_{R}^{c}\mathbf{\ h}_{0}$}

The Yukawa coupling for up quark fermions comes \ after the 2430$_{i}$ of E$%
_{6}$ take a vev. \ This scalar contains a SU(5) singlet with U(1) charges
(5,3) with respect U$_r$(1)$\times $U$_t$(1) and charge (-2,0) with respect U%
$_{T_3}$(1)$\times $U(1)$^\prime$. This belongs to the $(3,35)_{[ab]\rho
}^{\gamma }$ of \ SU(2)$\times $SU(6). The SU(5) singlet is in the direction 
$(3,35)_{[22]\rho }^{1i}$ , where we have also added the family index $i$ ,
in order to break the permutation symmetry and \ to induce a hierarchy \
among the top, the charm and the up quark. $\mathbf{T}_{(\alpha \beta )}$
stands for the fermions contained in the (1,15)\ of SU(2)$\times $SU(6). The
SU(6)\ invariant and completely antisymmetric tensor is $\ \varepsilon
^{\alpha \beta \delta \sigma \varepsilon \rho }.$ The operator responsible
for these masses is 
\begin{eqnarray*}
\mathbf{27}_{i}^{\alpha }\mathbf{\,\ 27}_{i}^{\beta }\,\ 351_{\gamma \delta
}^{\prime }\,2430_{\alpha \beta }^{i\,\ \gamma \delta } &=&\mathbf{T}%
_{(\alpha \beta )}\mathbf{T}_{(\delta \sigma )}(3,21)_{[\gamma \varepsilon
]}^{[ab]}(3,35)_{[ab]\rho }^{\gamma }\varepsilon ^{\alpha \beta \delta
\sigma \varepsilon \rho } \\
&=&\mathbf{T}_{(\alpha \beta )}\mathbf{T}_{(\delta \sigma
)}(3,21)_{[15]}^{[22]}(3,35)_{[22]\rho }^{1}\varepsilon ^{\alpha \beta
\delta \sigma 15}= \\
&=&\mathbf{u}_{L}^{i}\mathbf{u}_{R}^{ci}\ \ \ h_{0}\left(
(3,35)_{[22]1}^{1i}-(3,35)_{[22]5}^{5i}\right)
\end{eqnarray*}

\subsubsection*{$\mathbf{\bar{D}D+\bar{N}}_{L}\mathbf{N}_{L}\mathbf{+\bar{E}}%
_{L}\mathbf{E}_{L}$}

These masses are necessary and are expected to be very heavy, close to the
unification scale. Note that the fermions $D$ and $F$ belong to the same
SU(5) representation, so we have to be sure that the mass $T_{(1\beta
)}R_{1i}^{\beta }=\left( \bar{D}D+\bar{N}_{L}N_{L}+\bar{E}_{L}E_{L}\right) $
is much larger than $T_{(1\beta )}R_{2i}^{\beta }=\left( \bar{D}\,d_{R}^{c}+%
\bar{N}_{L}\nu_{L}+\bar{E}_{L}e_{L}\right) $ otherwise the lightest
neutrinos and down quarks would have not the desired U(1) charges. To to
that we need \ that the 27$_{22}$ \ , (i.e. the scalar component with
charges (0,4)) takes a vev much larger than the 22$_{16}$.

\begin{eqnarray*}
27_{i}^{\alpha }\,\ 27_{i}^{\beta }\,\ 27_{k\,}^{\gamma }\,\varepsilon
_{\alpha \beta \gamma } &=&T_{(\alpha \beta )}R_{bi}^{\beta }(2,\bar{6}%
)_{a}^{\alpha k}\varepsilon ^{ab}=T_{(1\beta )}R_{1i}^{\beta }(2,\bar{6}%
)_{2}^{1k}\varepsilon ^{12}= \\
&=&\left( \bar{D}D+\bar{N}_{L}N_{L}+\bar{E}_{L}E_{L}\right) (2,\bar{6}%
)_{2}^{1k}
\end{eqnarray*}

\subsubsection*{$\mathbf{\left( d_{R}^{c}d_{L}+e_{L}e_{R}^{c}\right)}
 \mathbf{h}_{0}$}

Finally we have the yukawa coupling for the down quarks and the charged
 leptons. They appear after
the 1728$_{\beta \delta }^{*\rho }$ takes a vev. This \ representation
contains a SU(5)\ singlet with charges (5,7) that is within the (4,\={6})$%
_{\alpha }^{[abc]}$ of SU(2)$\times $SU(6).\ Differently from the up quark
Yukawa coupling, the 1728 can accommodate not left-right symmetric mass
matrices. This \ is because \ we have the choice between ${1728^*}_{i\,\
\beta \delta }^{\rho }$ \ and $1728_{j\,\ \beta \delta }^{*\rho }$ in the
operator below. In other words the family index $i$ in the $%
(4,6)_{i1}^{[222]}$ can be attached either to the $d_{L}$ or the $d_{R}^{c}$
giving rise to distinct matrices. \ \ 

\begin{eqnarray*}
27_{i}^{\alpha }\,\ 27_{j}^{\beta }\,\ 351^{\prime \ast \,\ \gamma \delta
}\,\ {1728^*}_{i\,\ \beta \delta }^{\rho }\,\varepsilon _{\alpha \gamma \rho
} &=&T_{(\alpha \beta )i}R_{aj}^{\beta }\,(\bar{3},\bar{21})_{[bc]}^{[\alpha
\gamma ]}(\bar{4},{6})_{i}^{\alpha \lbrack abc]}= \\
&=&T_{(5\beta )i}R_{2j}^{\beta }\,(\bar{3},\bar{21})_{[22]}^{[51]}(\bar{4}%
,6)_{i1}^{[222]}= \\
&=&\left( x_{Lj}\,N_{Li}+d_{Rj}^{c}d_{Li}+e_{Lj}e_{Ri}^{c}\right) \,\
h_{0}\,\ (4,6)_{i1}^{[222]}
\end{eqnarray*}

\section*{Conclusion}

We have studied a grand unified model based on the E$_{6}$ unification
group. The three families are identical in the context of third
quantization, that means that the fundamental lagrangian is invariant under
permutations of the three fermion families. This permutation symmetry is
manifest in the neutrino sector, where large mixing angles together with \ s$%
_{13}\sim 0$ \ result, clearly point toward a soft S$_{3}$ symmetry breaking
pattern. In the quark sector, mixing angles are small and this clearly
restrict the class of grand unified theories that could be considered as
good candidate to describe the full spectrum of fermion masses. In fact the
first compelling issue is how to make a distinction between the up quarks
and the neutrino sector. In SO(10) we have an additional U(1) belonging to
the cartan algebra, and an additional right handed neutrino that is a
standard model singlet. From a simple analysis of the U(1) charges is not
obvious how to make a distinction between the up quark and the dirac
neutrino mass. The 16 16 10 invariant yukawa interaction obtained from the
product of two fermion families belonging to the 16 and the scalar
representation belonging to the 10 of SO(10), give the same yukawa both for
up quarks and neutrinos. Since we want Dirac neutrino mass matrix
 proportional to the identity
matrix we would obtain the
unacceptable $m_{\text{top}}=m_{\text{charm}}=m_{\text{up}}$ up quark
masses. If we enlarge the group to E$_{6}$, we have two additional
right-handed neutrino instead of one. This new neutrino can mix with the
light left-handed neutrino through a Yukawa operator that is distinct from
up quark yukawa. This can be achieved if we put the standard Higgs doublet
in the 351$^{\prime }$ with $U(1)_r\times U(1)_t$ charges equal to (+3,+5).
 We have studied the feasibility of this  scenario. At
the tree level there is only only one yukawa interaction, mixing the the
light left- handed neutrino and standard model singlet $\nu _{L}\,x_{L}h_{0}$%
. The remaining Yukawa coupling, necessary to turn on masses for all matter
fermions , arise only after loop correction are included, and appear as
higher order operators. We have found that is necessary to turn on the vev
of the 2430 or 1728, otherwise the top mass would be zero. From the tensor
analysis, we conclude this scenario  is feasible but further work, including
 radiative
corrections and an explicit analysis of the effective potential, is needed
to have a complete and exhaustive description of the model.

\end{document}